\documentclass[a4paper]{jpconf}
\usepackage{graphicx,url,amssymb,amsmath}

\newcommand{\td}[3]{\frac{d^{#3} #1}{d {#2}^{#3}}} 
\renewcommand{\v}[1]{\ensuremath{\mathbf{#1}}} 

\begin{document}
\title{Multi-messenger hunts for heavy WIMPs}

\author{Geoff Beck}

\address{School of Physics, University of the Witwatersrand, Private Bag 3, WITS-2050, Johannesburg, South Africa}

\ead{geoffrey.beck@wits.ac.za}

\begin{abstract}
Heavy neutrinos have a long history of consideration in the literature, in particular related to their role as solutions to the problems of neutrino mass, baryon asymmetry, and possibly dark matter. Interestingly, recent developments in the Madala hypothesis, a standard model extension designed to explain persistent LHC lepton anomalies, may also necessitate a heavy neutrino. This prospect is exciting as a dark matter model consisting of a TeV-scale leptophilic fermionic particle is also invoked to explain the electron-positron excess observed by the DAMPE experiment. The tantalising similarities between these new fermions may allow indirect dark matter detection methods to probe empirically compelling standard model extensions, like the Madala hypothesis. However, the leptophilic nature and large mass mean the expected gamma-ray signatures of annihilation or decay are weaker than those in the traditionally considered heavy quark and tau lepton channels. In this work we explore whether the KM3NeT neutrino detector could take advantage of the leptophilic nature of the added particle to provide an alternative means of exploring such interesting connections between cosmology and collider physics. We demonstrate that dwarf galaxies, in particular highly dense ultra-faint dwarf galaxies like Triangulum II, provide very strong prospects for KM3NeT searches.    
\end{abstract}

\section{Introduction}
\begin{figure}[!ht]
	\centering
	\resizebox{0.7\hsize}{!}{\includegraphics{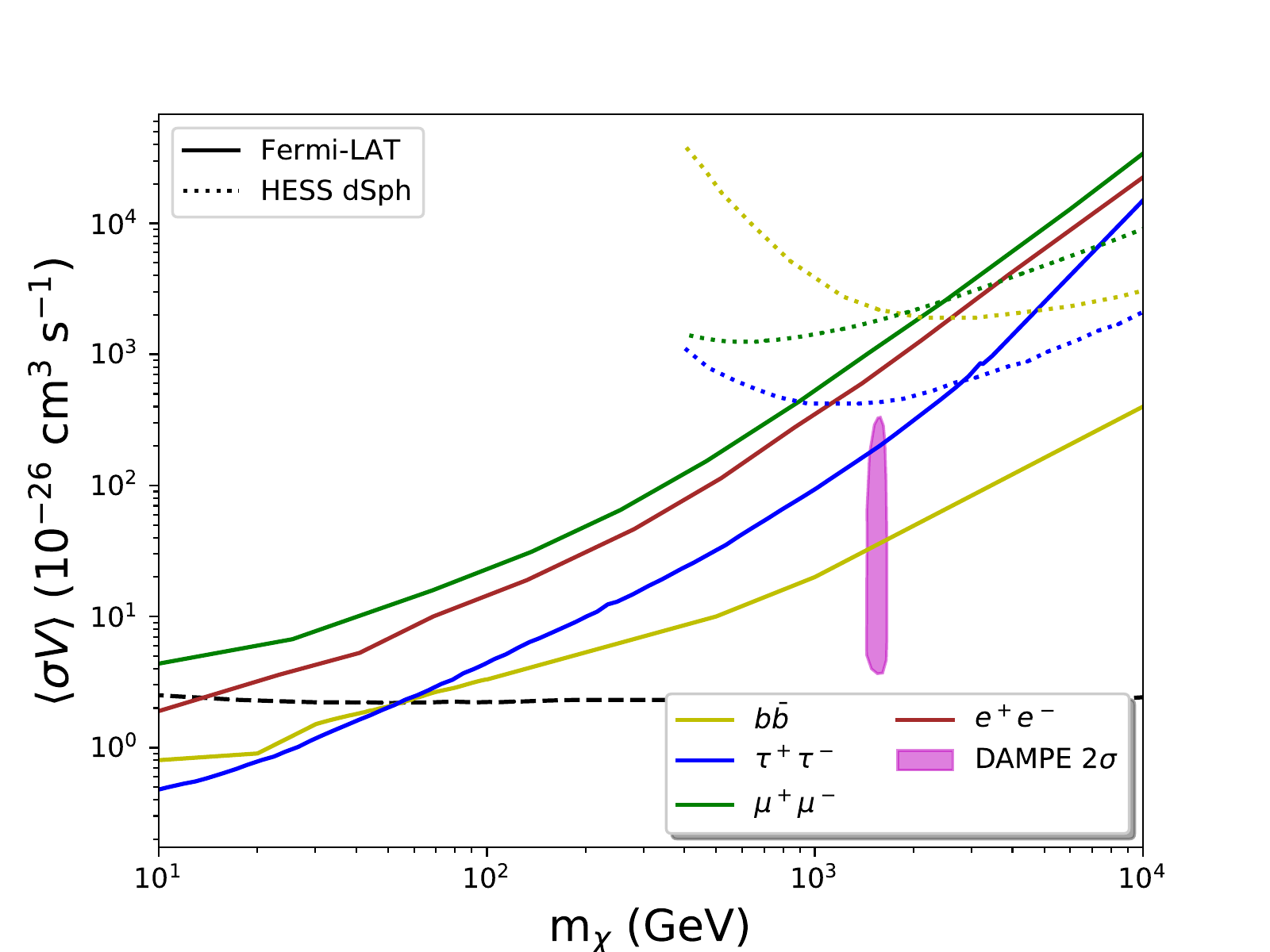}}
	\caption{Limits on the DM annihilation cross-section from Fermi-LAT~\cite{Fermidwarves2015,Fermidwarves2016} and HESS~\cite{hess-dsph-2014}. The contour from \cite{dampedm1} lies well below the muon/electron channel limits. The black dashed line shows the themal relic cross-section~\cite{steigman2012}.}
	\label{fig:current}
\end{figure}

Indirect searches for Dark Matter (DM) have managed to probe below the thermal relic cross-section for Weakly Interacting Massive Particles (WIMPs) with masses below 100 GeV in a variety of environments and frequency ranges~\cite{Fermidwarves2015,Fermidwarves2016,egorov2013,beckm312019}. High energy gamma-rays may offer a promising future probe via the up-coming Cherenkov Telescope Array but DM particles that annihilate via muons and electrons in particular produce significantly weaker cross-section constraints in gamma-rays~\cite{Fermidwarves2015} (as can be seen in Figure~\ref{fig:current}). Models that couple DM to lighter leptons are of particular interest on two grounds: the first being that it has been shown that the DArk Matter Particle Explorer (DAMPE) excess~\cite{dampe} can be explained in terms of these models~\cite{dampedm1,dampedm2,dampedm3,dampeucmh}, the second being the persistent lepton anomalies emerging in Large Hadron Collider (LHC)~\cite{atlas-docs,cms-docs} data as discussed in \cite{madala1,madala3,madala-v2-2}. This second point is linked to the Madala hypothesis, an extension of the standard model designed to explain LHC anomalies that have grown more significant as more data has emerged (see \cite{madala1,madala2,madala3,madala4} for the original hypothesis). The hypothesis originally contained a coupling to a dark sector, which was used to constrain the properties of additional particles in \cite{gsmadala1,gsmadala2,gsmadala3}. However, a necessary reformulation has moved the Madala scenario away from an effective field theory approach~\cite{madala-v2-1,madala-v2-2}. This alteration has also removed the necessity of the original dark sector coupling. However, it does include room for, and may even require, an extra heavy neutrino~\cite{lhc-leptons-madala}, which may be employed to resolve the neutrino mass problem~\cite{drewes2015}, as well as both the muon anomalous magnetic moment~\cite{muong-2,muong-2-exp} and anti-baryon asymmetry~\cite{drewes2015}. Crucially, the mass of this added neutrino is not well constrained, with \cite{lhc-leptons-madala} considering $100$ to $1000$ GeV. In general, \cite{drewes2015} notes that the mass could be anywhere between the eV and GUT scales but that TeV scale neutrinos are of significant interest phenomenologically (see for instance \cite{akhmedov2013}). Interestingly, leptophilic WIMP models proposed to account for the DAMPE excess can also provide a mechanism for producing neutrino masses~\cite{dampedm1} and typically invoke a fermionic WIMP with a mass around 1 TeV.

In this work we explore the idea that a neutrino-like particle with a mass around 1 TeV could explain all the aforementioned anomalies. We will do this by considering the potential to detect annihilation of the particle indirectly through up-coming neutrino telescopes such as KM3NeT. We particularly consider annihilation here in order to be relevant to the cross-section parameter space for the DAMPE excess from \cite{dampedm1}.

In previous works \cite{gsdampe1,gsdampe2} we explored the potential of various approaches to detecting local DM over-densities associated with the DAMPE excess \cite{dampe,dampedm1,dampedm2,dampedm3,dampedm4,dampeucmh} in radio, gamma-rays, and neutrinos. In so doing we found that a promising method of constraint on leptophilic WIMPs would be to employ the KM3NeT detector to search for a neutrino flux from the galactic centre (despite the large angular extension). Since we are looking for heavy WIMPs our aims in this work overlap somewhat with our aforementioned works. However, we will extend the preceding results by looking at the use of neutrino observations to constrain the WIMP parameter space in a larger mass range around $1$ TeV. In particular, we will focus upon the potential of dwarf galaxies as neutrino observation targets. This will allow us to both find new environments from which to probe the DAMPE excess models of DM, but will also allow us to investigate the potential of Madala hypothesis associated particles to solve the DM problem. Thus, this work will serve to expand our previous approach \cite{gsmadala1,gsmadala2,gsmadala3} to the reformulation of the Madala hypothesis that no longer relies upon an effective field theory~\cite{madala-v2-1,madala-v2-2}.

We demonstrate that, in addition to the galactic centre~\cite{gsdampe2}, dwarf spheroidal galaxies have strong potential to improve probes on leptophilic WIMP models with masses around $1$ TeV. In particular, despite the uncertainties over its J-factor, Triangulum II provides a very prominent candidate for neutrino based searches. This is because, even in the most pessimistic case, it can provide improvements on the constraints obtained by Fermi-LAT, and, in the most optimistic cases, probe the entire DAMPE excess parameter space and encroach upon the thermal relic cross-section. This would allow for significant exploration of the further consequences of the Madala hypothesis, should it be able to accommodate a neutrino-like particle suitable as a DM candidate, and thus determine whether DM and LHC leptonic anomalies can be connected. 

This work is structured as follows: in Section~\ref{sec:emm} we discuss the estimation of neutrino fluxes from DM annihilation, while the target halo properties and KM3NeT sensitivities are discussed in Sections~\ref{sec:halos} and \ref{sec:km3net} respectively. Finally, results are presented in Section~\ref{sec:results} and discussed in \ref{sec:conc}. 

\section{Dark matter emmission of neutrinos}
\label{sec:emm}
The flux of neutrinos from DM annihilation can be written in terms of the $J$ factor
\begin{equation}
J (\Delta \Omega, l) = \int_{\Delta \Omega}\int_{l} \rho^2 (\v{r}^{\prime}) dl^{\prime}d\Omega^{\prime} \; , \label{eq:jfactor}
\end{equation}
where $l$ is the line of sight and $\Delta \Omega$ is the angular extension of the target.
The flux is then found via
\begin{equation}
S_{\nu} (E) = \frac{1}{2}\langle \sigma V\rangle \sum\limits_{f}^{} \td{N^f_\nu}{E}{} B_f J(\Delta \Omega,l) \; ,
\end{equation}
where $\langle \sigma V\rangle$ is the velocity averaged annihilation cross-section, $\td{N^f_\nu}{E}{}$ is the number of neutrinos produced per annihilation per unit energy via annihilation channel $f$, $B_f$ is the branching fraction of channel $f$, and $\rho(\chi)/m_{\chi}$ is the WIMP number density. 

We will follow the standard approach where we set $B_f = 1$ and study each of the tau, muon, and electron leptonic channels individually. The functions $\td{N^f_i}{E}{}$ will be sourced from \cite{ppdmcb1,ppdmcb2} including electro-weak corrections.

\section{Target halos}
\label{sec:halos}

In this work we focus upon dwarf galaxies and so present the J-factors for the Draco, Reticulum II, Triangulum II dwarf galaxies. These are chosen as they have all been considered as DM hunting environments in the past~\cite{Colafrancesco2007,bonnivard2015,regis2017}, with recent attention on the case of Triangulum II in particular~\cite{genina2016}.

\begin{table}[!ht]
	\begin{tabular}{|l|l|l|l|l|l|}
		\hline
		Halo & J-factor (min) & J-factor (median) & J-factor (max) & $\theta$ & Reference\\
		\hline
		Triangulum II & $1.86 \times 10^{19}$ & $2.75\times 10^{20}$ & $4.37\times 10^{21}$ & $0.5^{\circ}$ & \cite{genina2016,Hayashi:2016kcy,pace2019}\\
		Reticulum II & $7.24\times 10^{16}$ & $5.75\times 10^{17}$ & $4.90\times 10^{18}$ & $0.5^{\circ}$ & \cite{bonnivard2015,Hayashi:2016kcy,pace2019} \\
		Draco & $5.37\times 10^{18}$ & $1.23\times 10^{19}$ & $3.02\times 10^{19}$ & $0.5^{\circ}$ &\cite{Hayashi:2016kcy,pace2019} \\
		\hline
	\end{tabular}
\caption{A summary of the J-factor values used in this work. $\theta$ refers to angular radius over which $J$ is calculated. All J-factors are given in units of GeV$^2$ cm$^{-5}$.}
\label{tab:halos}
\end{table}

\section{KM3NeT sensitivities}
\label{sec:km3net}
We follow \cite{Ambrogi:2018skq} in using their calculations for the sensitivity of KM3NeT to extended sources. Note that these only consider muon neutrinos while KM3NeT is expected to be sensitive to all three flavours. To account for this we will consider two cases: a pessimistic one where we only compare predicted DM muon neutrino fluxes to the sensitivity from \cite{Ambrogi:2018skq}, and an optimistic one where we use a DM flux with all three flavours. These two extremes should serve to bracket the expected range of non-detection constraints that could be achieved.

\section{Results}
\label{sec:results}
Here we present the non-detection constraints for the 3 dwarf galaxies considered. These are displayed relative to existing Fermi-LAT results from \cite{Fermidwarves2015,Fermidwarves2016} with the thermal relic value from \cite{steigman2012} for comparison. We will display results for annihilation to all lepton channels as we are interested in leptophilic models of DM. However, we note the muon and electron channels are of special interest due to their role in models suggested to explain the DAMPE excess~\cite{dampedm1} and the significance of these leptons in LHC excesses~\cite{madala-v2-2}.

\begin{figure}[!ht]
	\centering
	\resizebox{0.45\hsize}{!}{\includegraphics{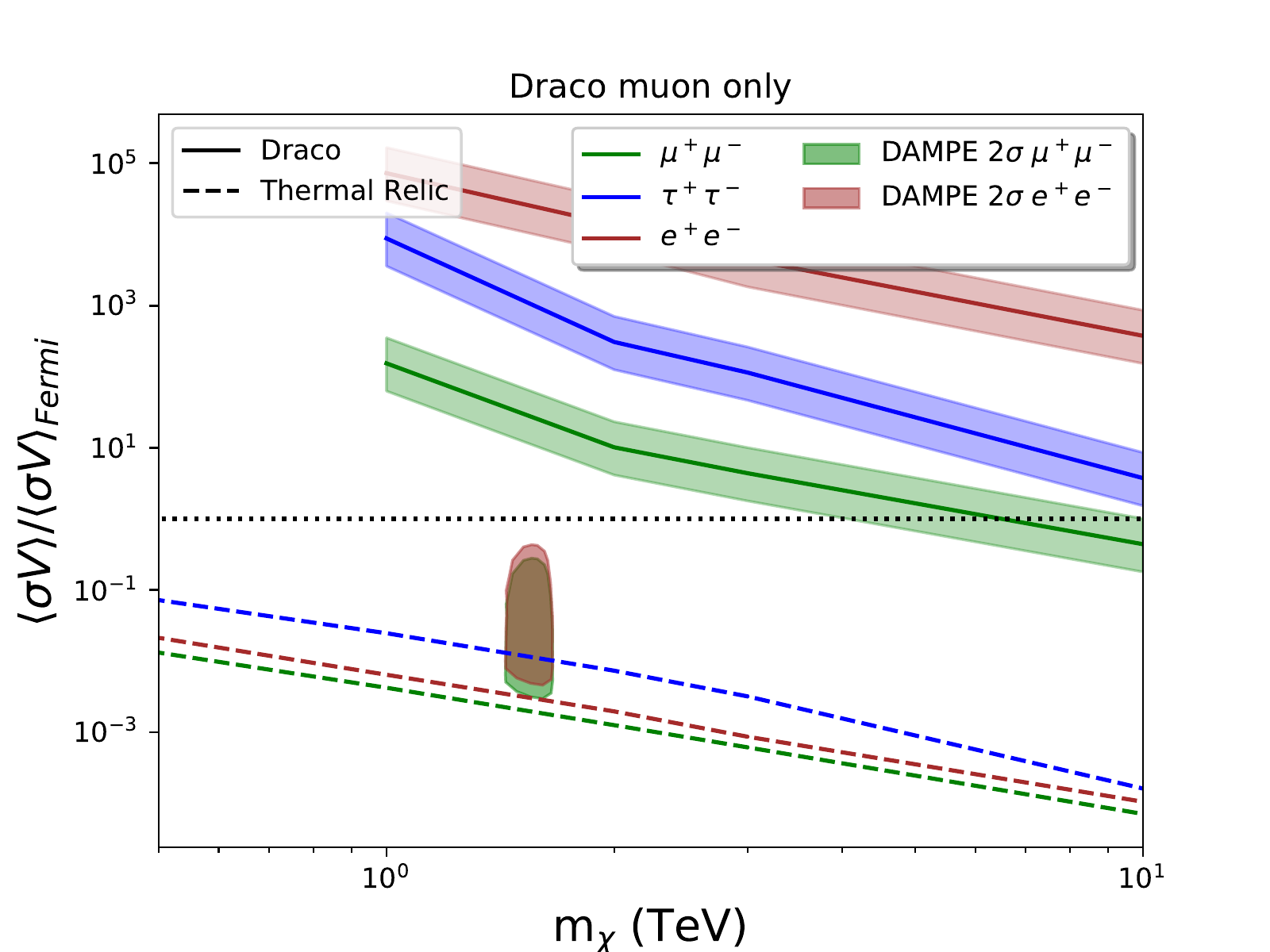}}
	\resizebox{0.45\hsize}{!}{\includegraphics{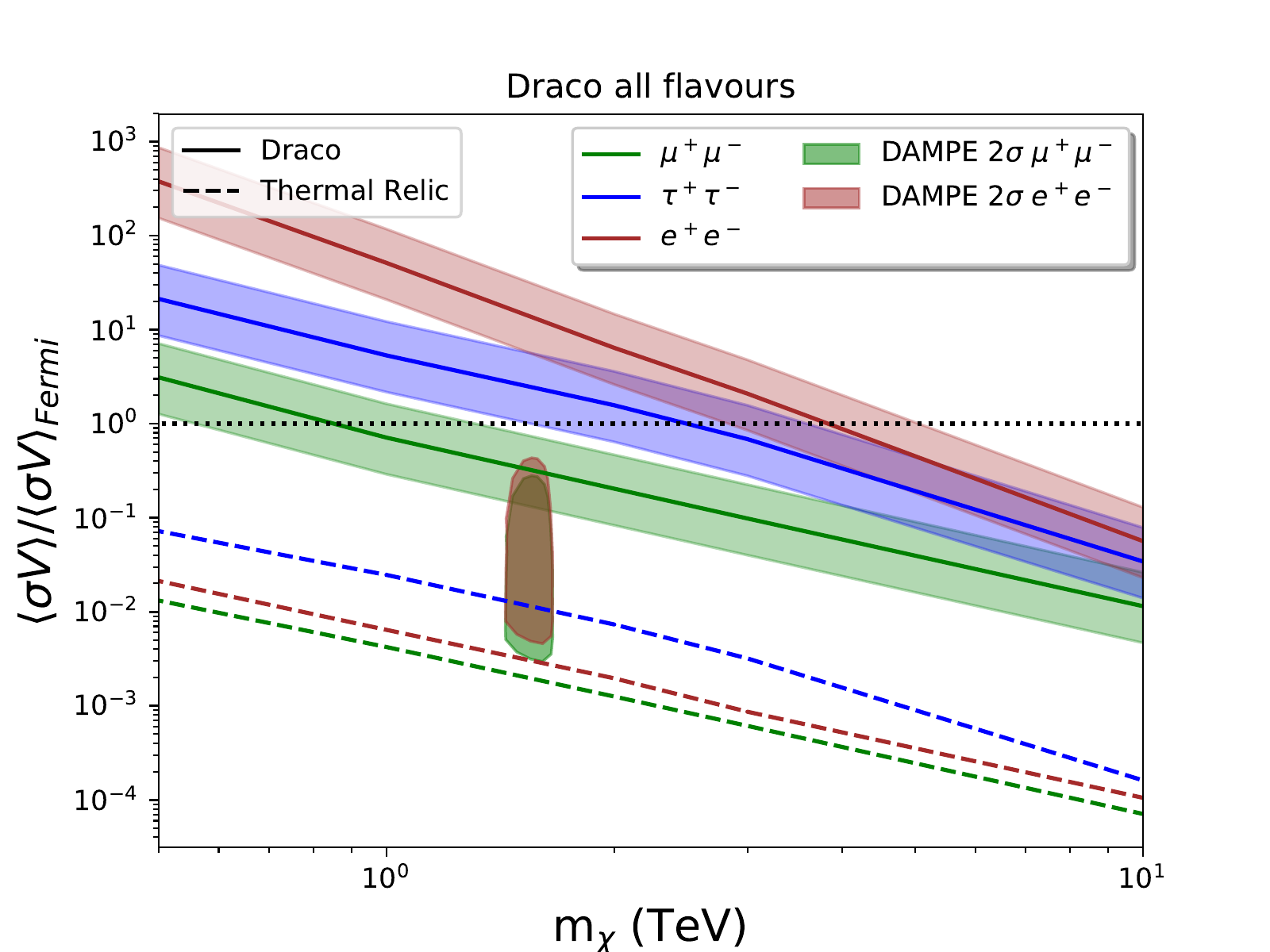}}
	\caption{Non-observation upper limits on the annihilation cross-section from the Draco dwarf with KM3NeT. The solid line shows the median J-factor case while shading is between the minimum and maximum values from Table~\ref{tab:halos}. Left: muon neutrinos only. Right: all flavours.}
	\label{fig:draco}
\end{figure}

In Figure~\ref{fig:draco} we display the results for the Draco dwarf. Even in the most optimistic case this target can only marginally constrain the DAMPE parameter space via KM3NeT non-observation in the muon channel. However, it produces improvements over Fermi-LAT for $m_{\chi} >  7$ TeV in the pessimistic case, and $m_{\chi} > 1$ TeV in the optimistic one (for tau and electon channels this occurs at masses $3$ or $4$ times larger). Notably, the constraints improve with WIMP mass (at least in the displayed range), unlike those from other high-energy measures like gamma-rays. Importantly, the muon neutrino only case significantly worsens the limit on the electron channel as expected.

\begin{figure}[!ht]
	\centering
	\resizebox{0.45\hsize}{!}{\includegraphics{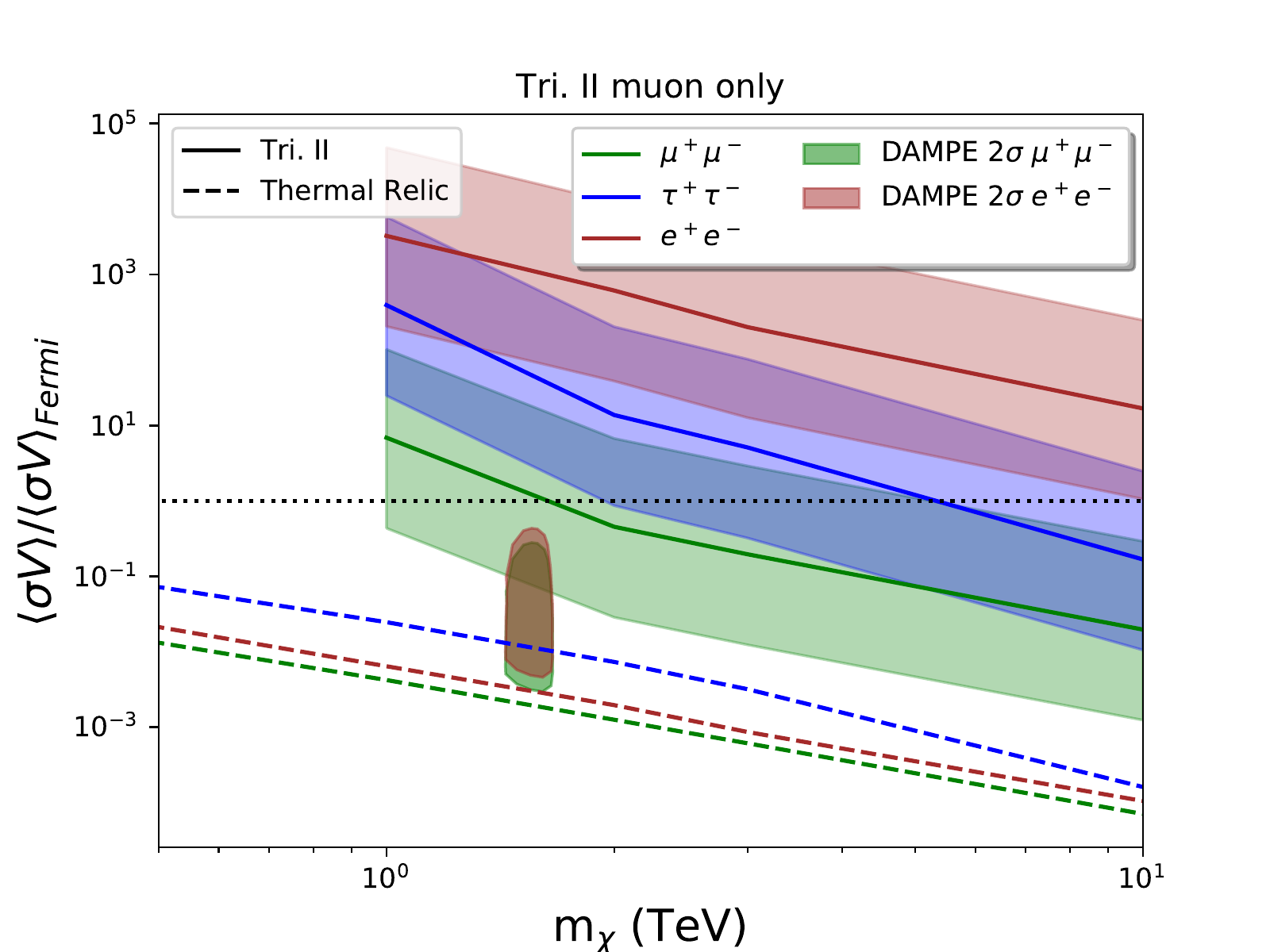}}
	\resizebox{0.45\hsize}{!}{\includegraphics{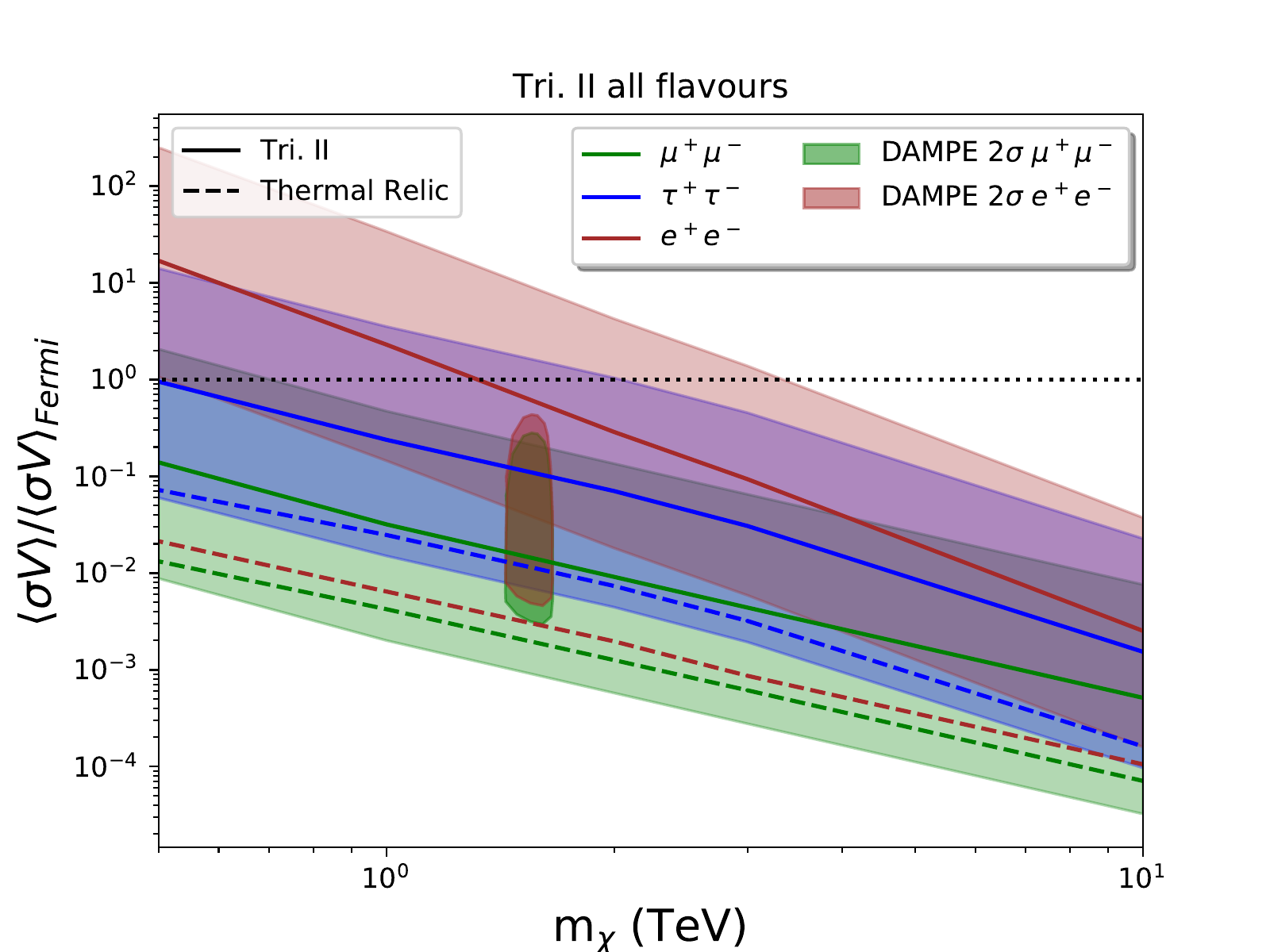}}
	\caption{Non-observation upper limits on the annihilation cross-section from the Triangulum II dwarf with KM3NeT. The solid line shows the median J-factor case while shading is between the minimum and maximum values from Table~\ref{tab:halos}. Left: muon neutrinos only. Right: all flavours.}
	\label{fig:tri2}
\end{figure}

Figure~\ref{fig:tri2} displays the case of the Triangulum II dwarf. This case has substantial uncertainties in the J-factor~\cite{genina2016,Hayashi:2016kcy,pace2019} but manages to improve upon Fermi-LAT substantially in both scenarios for tau and muon channels. Importantly, the optimistic case achieves significant coverage of the DAMPE parameter space and encroaches upon the thermal relic density.

\begin{figure}[!ht]
	\centering
	\resizebox{0.45\hsize}{!}{\includegraphics{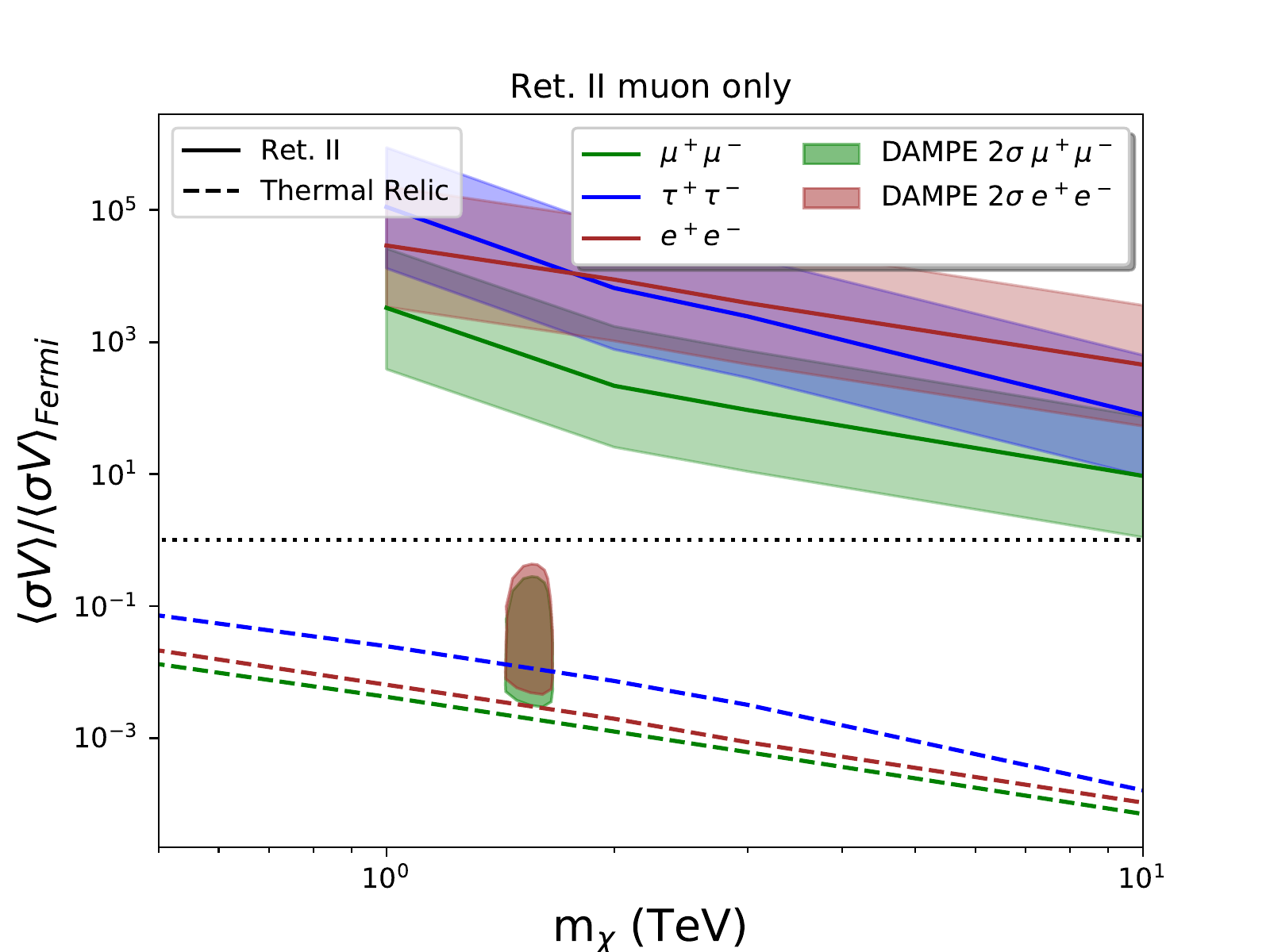}}
	\resizebox{0.45\hsize}{!}{\includegraphics{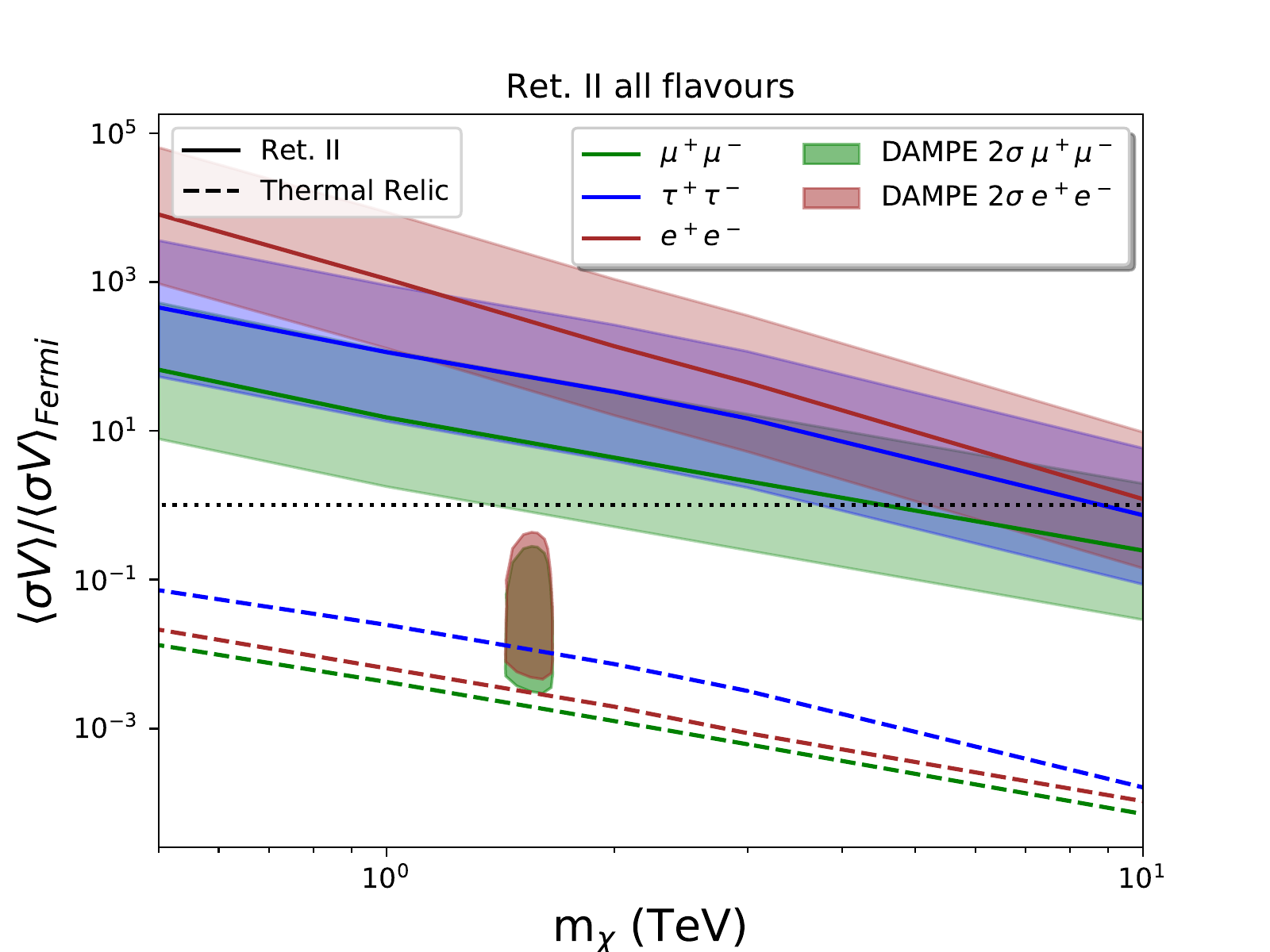}}
	\caption{Non-observation upper limits on the annihilation cross-section from the Reticulum II dwarf with KM3NeT. The solid line shows the median J-factor case while shading is between the minimum and maximum values from Table~\ref{tab:halos}. Left: muon neutrinos only. Right: all flavours.}
	\label{fig:ret2}
\end{figure}

Reticulum II, displayed in Fig.~\ref{fig:ret2} fails to encroach upon the DAMPE parameter space in either case and even at the most optimistic J factors. However, in the optimistic case, it does manage to improve upon Fermi-LAT for $m_{\chi} > 1.5$ TeV and $m_{\chi} > 5$ TeV for the maximum and median J factors respectively.

\section{Discussion and conclusions}
\label{sec:conc}
The WIMP mass range $> 1$ TeV is a difficult regime to probe. However, the results presented here demonstrate that substantial advances can be made in leptonic annihilation channels (specifically relevant to leptophilic DM models) through the use of non-detection constraints from the up-coming KM3NeT experiment. It is especially notable that the cross-section constraints derived in this manner improve with increasing WIMP mass, in sharp contrast to the behaviour of both radio and gamma-ray limits. These particular annihilation channels are of special interest as they offer a chance to probe models of DM that may have relevance to multiple outstanding issues in modern physics and provide an alternative means of effectively probing standard model extensions, such as the Madala hypothesis, which provide potential leptophilic DM candidates. In future work it may become possible to combine LHC and astrophysical/cosmological constraints to determine if a leptophilic standard model extension can compellingly address anomalies across such vastly different scales. 

\section*{Acknowledgments}
G.B acknowledges support from a National Research Foundation of South Africa Thuthuka grant no. 117969. 

\section*{References}
\bibliographystyle{iopart-num}
\bibliography{heavy_wimps_madala.bib}

\end{document}